# Weyl-fermions, Fermi-arcs, and minority-spin carriers in ferromagnetic CoS$_2$

Niels B. M. Schröter[1*†], Iñigo Robredo[2,3*], Sebastian Klemenz[4], Robert J. Kirby[4], Jonas A. Krieger[1,5,6], Ding Pei[7], Tianlun Yu[1], Samuel Stolz[8,9], Thorsten Schmitt[1], Pavel Dudin[10], Timur K. Kim[10], Cephise Cacho[10], Andreas Schnyder[11], Aitor Bergara[2,3,12], Vladimir N. Strocov[1], Fernando de Juan[2,13], Maia G. Vergniory[2,13†], Leslie M. Schoop[3†].

\* contributed equally
† correspondence to niels.schroeter@psi.ch, maiagvergniory@dipc.org, lschoop@princeton.edu

[1]Swiss Light Source, Paul Scherrer Institute, CH-5232 Villigen PSI, Switzerland
[2]Donostia International Physics Center, 20018 Donostia-San Sebastian, Spain
[3]Condensed Matter Physics Department, University of the Basque Country UPV/EHU, 48080 Bilbao, Spain
[4]Department of Chemistry, Princeton University, Princeton, New Jersey 08540, USA
[5]Laboratorium für Festkörperphysik, ETH Zurich, CH-8093 Zurich, Switzerland
[6]Laboratory for Muon Spin Spectroscopy, Paul Scherrer Institute, CH-5232 Villigen PSI, Switzerland
[7]Clarendon Laboratory, Department of Physics, University of Oxford, Oxford OX1 3PU, United Kingdom
[8]EMPA, Swiss Federal Laboratories for Materials Science and Technology, 8600 Dübendorf, Switzerland
[9]Institute of Condensed Matter Physics, Station 3, EPFL, 1015 Lausanne, Switzerland
[10]Diamond Light Source, Didcot, OX110DE, United Kingdom
[11]Max Planck Institute for Solid State Research, 70569, Stuttgart, Germany
[12]Centro de Física de Materiales, Centro Mixto CSIC -UPV/EHU, 20018 Donostia, Spain
[13]IKERBASQUE, Basque Foundation for Science, Maria Diaz de Haro 3, 48013 Bilbao, Spain

## Abstract

The pyrite compound CoS$_2$ has been intensively studied in the past due to its itinerant ferromagnetism and potential for half-metallicity, which make it a promising material for spintronic applications. However, its electronic structure remains only poorly understood. Here we use complementary bulk- and surface-sensitive angle-resolved photoelectron spectroscopy and ab-initio calculations to provide a complete picture of its band structure. We discover Weyl-cones at the Fermi-level, which presents CoS$_2$ in a new light as a rare member of the recently discovered class of magnetic topological metals. We directly observe the topological Fermi-arc surface states that link the Weyl-nodes, which will influence the performance of CoS$_2$ as a spin-injector by modifying its spin-polarization at interfaces. Additionally, we are for the first time able to directly observe a minority-spin bulk electron pocket in the corner of the Brillouin zone, which proves that CoS$_2$ cannot be a true half-metal. Beyond settling the longstanding debate about half-metallicity in CoS$_2$, our results provide a prime example of how the topology of magnetic materials can affect their use in spintronic applications.

# Main text

Since the experimental discovery of topological insulators and topological semimetals, there has been a substantial effort to functionalise novel topological materials for spintronic applications, most prominently in the family of Bismuth-chalcogenides[1–4]. However, because the field of spintronics predates the first prediction of topological insulators[5,6] and topological semimetals[7], topological phenomena in many well-established spintronic materials may have been overlooked during initial investigations of their electronic properties. Here we reveal that $CoS_2$, a material that has been long studied due to its itinerant ferromagnetism and potential for half-metallicity, actually hosts Weyl-fermions and Fermi-arc surface states in its band structure close to the Fermi-level $E_F$, as well as topological nodal-lines below the Fermi-level. $CoS_2$ is, therefore, a rare example of the recently discovered class of experimentally verified magnetic topological metals[8–10], which are of practical importance for device applications[11], and of broader interest for fundamental science, for instance as a platform to realize axion insulators[7], the intrinsic anomalous Hall effect[12], or the anomalous fractional quantum Hall effect[13].

Beyond the discovery of its topological properties, we also clarify the question of whether $CoS_2$ is a true half-metal. Half-metallic ferromagnets, i.e. materials that are metallic in one spin-channel but gapped in the other, are important components of spintronic devices where they act as sources of spin-polarized charge carriers. There has been a longstanding debate on whether $CoS_2$ or the related alloys $Co_{1-x}Fe_xS_2$ are true half metals, which has important implications for materials and device engineering. Calculations within the local spin-density approximation (LSDA)[14,15] (see Fig. 1A-B) conclude that $CoS_2$ hosts a minority-spin electron pocket at the R-point that leads to a peak in the minority-spin density of states (DOS), which would make $CoS_2$ a minority-spin conductor. This scenario is supported by magnetotransport experiments that suggest a sign flip in the spin-polarization upon hole-doping with iron that may turn $CoS_2$ into a half-metal (i.e. vanishing minority-spin DOS at $E_F$) due to the depopulation of a minority-spin electron pocket [16–20]. On the other hand, calculations based on the generalized gradient approximation (GGA)[15,21] (see Fig. 1B-C) conclude that half-metallicity is already obtained in the undoped compound, and quantum oscillation experiments were unable to detect the putative minority-spin electron pocket[21]. To resolve the debate about the half-metallicity of $CoS_2$, a direct spectroscopic band-structure investigation would be clearly desirable, also because the spin-polarization measured with transport probes can be reduced by surface defects[22] and can therefore not entirely rule out intrinsic half-metallicity.

Besides the bulk band structure, electronic surface states can also influence the spin-polarization at the Fermi-level, which becomes particularly important for heterostructure interfaces in spintronic devices where half-metals function as spin-injectors. One important class of surface states are topological Fermi-arcs in Weyl-semimetals, which are considered to be particularly robust against passivation due to the protection by topological bulk invariants. Our LSDA calculations of $CoS_2$ shown in Fig. 1A-B predict Weyl-nodes close to the Fermi-level on the $k_x=0$ and $k_y=0$ planes parallel to the magnetisation direction (see Fig. 1D), which should give rise to topological Fermi-arc surface states. Additionally, we find topological nodal lines a few hundreds of meV below the Fermi-level (also see supplementary). Since the concept of topological semimetals and Fermi-arc surface states was not yet established at the time, previous theoretical studies of $CoS_2$ overlooked these important features in its band-structure, which can have a decisive influence on spin-transport properties.

Previous angle-resolved photoelectron spectroscopy (ARPES) experiments[23–25] on $CoS_2$ were performed with photon energies between hv=20-120 eV where the inelastic mean free path of the photoelectrons can be expected to be close to its minimum ~0.5 – 1 nm, resulting in very surface sensitive probes. This implies that the signal from bulk bands is substantially broadened along the momentum direction $k_z$ perpendicular to the sample surface[26], which makes the identification of any bulk band dispersion in these experiments challenging, and no minority-spin bulk electron pocket in the vicinity of the R-point could be resolved. While previous studies did report the presence of surface states[25], it remained unclear whether these surface states are crossing the Fermi-level, and whether they are of topological origin.

Here we overcome these limitations by using complementary bulk-sensitive soft X-ray ARPES (hv > 400 eV) and surface-sensitive VUV-ARPES (hv < 120 eV) to disentangle the bulk and surface electronic structure of $CoS_2$. As a result, we are for the first time able to observe its elusive minority-spin bulk electron pocket at the R-point of the Brillouin-zone directly, settling the longstanding debate of half-metallicity in $CoS_2$. We also detect a topological line-node located at ~150 meV below the Fermi level, as well as a Weyl-cone that suggest a Weyl-point slightly above the Fermi-level, which gives rise to topological Fermi-arc surface states that cross the Fermi-level. By comparison with ab-initio calculations, we find that these Fermi-arcs are of majority spin-character and therefore affect the spin-polarization at a heterostructure interface.

Our samples were synthesized using chemical vapour transport. We studied their elemental composition with core-level spectroscopy and energy dispersive X-ray spectroscopy, which confirmed the expected stoichiometry (see Fig 1E). Powder x-ray diffraction confirmed the previously reported cubic space group 205 and a lattice constant of a= 5.5287(5) (see methods and supplementary for more details about synthesis and characterization). We measured a Curie temperature of Tc=124 K and a saturation magnetisation of 0.91 $\mu_B$/Co at 9 T (which becomes linear at 0.2 T and 0.89 $\mu_B$), in good agreement with the literature values[27] (see Fig. 1F for magnetisation curves).

When cleaving $CoS_2$ for ARPES experiments, we obtained two distinct cleavage planes with the surface normal pointing along the (111) and (100) directions. We performed photon energy-dependent ARPES measurements with soft X-ray photons (hv=350-800 eV) to locate the high-symmetry planes along the $k_z$ direction normal to the sample surface (see supplementary materials). The band structure in the $k_z=\pi$ plane containing the R-point is displayed in Fig. 2. For the data measured on the (111) cleavage plane, we can clearly identify circular Fermi-surface pockets at the R-point in the corner of the Brillouin zone. Our calculated Fermi-surface (Fig. 2B, employing the LSDA) is in good qualitative agreement with the experimental data, confirming the existence of Fermi-surface pockets at the R-point. When inspecting the experimental band dispersion along the R-X-R direction (Fig. 2C), we see that the circular pockets at the R-point are electron-like, and are related to another parabolic band with a minimum at around ~0.65 eV by the exchange splitting. The magnitude of the exchange splitting extracted from the energy distribution curve at the R point (Fig. 2D) is ΔE= 0.60(3) eV. Our LSDA calculations of the band dispersion shown in Fig. 2E are in good qualitative agreement with the experimental data and indicate that the observed electron pocket is of minority-spin character, which implies that $CoS_2$ is not a true half-metal. However, the experimentally observed exchange splitting is ~250 meV smaller than in the LSDA calculations, such that the majority spin-bands are located closer to the Fermi-level in the experiment than expected from the calculations. The data measured on the (100) surface also show electron pockets at the R-point and is displayed in the supplementary materials.

To search for the topological nodal-line and Weyl-nodes in $CoS_2$, we also probed the bulk band structure in the $k_z=0$ plane, containing the Γ-point, as illustrated in Figure 3. Fig 3A-D display the experimental and calculated Fermi-surfaces for the (111) and (100) cleavage planes, which are in good qualitative agreement. Fig. 3E shows the band dispersion along the M- Γ direction (black arrow in Fig. 3A) measured on the (111) surface (black arrow in Fig. 3A), and the Γ -M direction, measured on the (100) surface (black arrow in Fig. 3C). Note that the observable bands along these two directions are very different, possibly due to matrix element effects. The line cut obtained from the (111) surface shows a V-shaped feature centred at the M point, and a quasi-parabolic band centred at the Γ-point. In contrast, the dispersion obtained from the (100) surface shows a single band dispersing in the opposite direction from the quasi-parabolic band. To enhance the contrast of our data, we also show the corresponding second derivative spectrum in Fig. 3F. The combined band dispersion from both surfaces is illustrated in Fig. 3G, which displays the peak positions from a fit of the momentum distribution curves (MDCs). By comparison with the calculated band dispersion shown in Fig. 3H, we can see that the band crossing between the blue and red bands (from the (111) surface and (100) surface, respectively) is part of a topological line-node, while the blue bands form a Weyl-cone. The Weyl-point that corresponds to the Weyl-cone is shown in Fig. 3I, which displays the calculated band dispersion along the M*-Γ direction, where M* $(0,0.5,0.4581)\frac{2\pi}{a}$ is a point that is slightly displaced from M $(0,0.5,0.5)\frac{2\pi}{a}$ (see Fig. 1D), which is identical to the M-point within the experimental uncertainty. Since we cannot observe the band top of the blue bands along the M- Γ direction in our experimental data, we conclude that the Weyl-point must be located slightly above the Fermi-level.

Fermi-arc surface states are a hallmark of Weyl-points in topological semimetals. Therefore, the Weyl-points in $CoS_2$ must be accompanied by Fermi-arc surface states that are connecting the projections of the Weyl-points in the surface Brillouin zone. We used surface-sensitive VUV-ARPES to investigate the surface electronic structure of the (100) surface in $CoS_2$, the results of which are displayed in Fig. 4. Figure 4A shows the experimentally obtained Fermi-surface, which was measured on a strongly tilted crystal plane. The photon energy dependence of the Fermi-surface maps (see supplementary materials) indicates that all Fermi-surface pockets measured by VUV-ARPES are surface states and that the signal from bulk states is mostly suppressed in this photon energy range. The Weyl-points identified in our ab-initio calculations project into the $\bar{X}-\bar{\Gamma}-\bar{X}$ path and the corresponding line-cut of our experimental data is shown in Fig. 4B. It displays a surface state band crossing the Fermi-level that connects two hole-like pockets that are located at the $\bar{X}$ points at opposite ends of the Brillouin zone. Our ab-initio calculations of the surface electronic structure displayed in Fig. 4C indicate that this surface state is a Fermi-arc that connects two Weyl-points. Note that the renormalized energy scale of the Fermi-arcs in the calculation compared to the experiment is expected due to the reduced exchange splitting, which we already observed for the bulk band structure. Therefore, the Fermi-arcs cross the Fermi-level in the experiment, while they are located below the Fermi-level in the calculation. Fig. 4D-E show that same line-cut measured at different photon energies, which shows that the Fermi-arc dispersion is independent of the $k_z$ momentum, which proves its two-dimensional nature. Since the Fermi-arcs are derived from majority-spin bulk bands, they have majority-spin character. This will influence the spin-polarization of electrons at the Fermi-level at heterostructure interfaces where $CoS_2$ can act as a spin-injector.

Due to the relatively low effective mass of the minority-spin electron pocket at the R-point, the minority spin-density of states at the Fermi-level is large, and some calculations suggested that $CoS_2$ is a minority-spin conductor[18]. Our discovery of the majority-spin Fermi-arc surface implies that the total interface spin-polarization of a heterostructure involving $CoS_2$ will be reduced compared to the bulk value. Compared to trivial dangling bond surface states, Fermi-arc surface states are more robust against attempts of passivation, because they are protected by the topological invariants (the Chern numbers) of the bulk Weyl points; hence engineering of the interface potential cannot easily remove the Fermi-arcs. On the other hand, hole doping with iron has been suggested to transform $CoS_2$ from a minority-spin to a majority-spin conductor and ultimately to a full half-metal by depopulation of the minority-spin bulk electron pocket. Such a transition to half-metallicity will be facilitated by the majority-spin Fermi-arcs at interfaces in heterostructures since they compensate the DOS of the minority-spin electron pocket. Therefore, $CoS_2$ provides a prime example of a spintronic material whose performance is affected by its topologically nontrivial band structure. There may be magnetic domains on the sample surface that are smaller than the size of our beamspot (dia. 50-70 µm), but based on our ab-initio calculations, we believe that the direction of the magnetization vector will have a negligible effect on the interpretation of the Fermi-arc structure (see supplementary for further discussion). $CoS_2$ is also known as a good catalyst, e.g. for the hydrogen evolution reaction[28], and it has recently been speculated that Fermi-arcs in Pt- and Pd-based chiral topological semimetals[29,30] may play a role in catalysis due to their d-electron character and their robustness against hydrogen passivation[31,32]. Since the Fermi-arcs in $CoS_2$ are also derived from bulk bands of d-orbital character (see supplementary), they may contribute to the catalytic performance of $CoS_2$.

# Acknowledgements

The authors would like to thank Igor Mazin, Matthew Watson, and Fabio Orlandi for fruitful discussions and helpful feedback. This work was supported by NSF through the Princeton Center for Complex Materials, a Materials Research Science and Engineering Center DMR-1420541, by Princeton University through the Princeton Catalysis Initiative, and by the Gordon and Betty Moore Foundation through Grant GBMF9064 to L.M.S. F. J. acknowledges funding from the Spanish MCI/AEI through grant No. PGC2018-101988-B-C21. M. G. V. and F. J. acknowledge funding from the Basque Government through grant PIBA 2019-81. M.G.V. thanks support from DFG INCIEN2019-000356 from Gipuzkoako Foru Aldundia. N.B.M.S. was supported by Microsoft. The work of S.S. was supported by the Swiss National Science Foundation under Grant No. 159690. D.P. acknowledges the support from Chinese Scholarship Council. J.A.K. acknowledges support by the Swiss National Science Foundation (SNF-Grant No. 200021_165910).

# References

1. Fan, Y. *et al.* Magnetization switching through giant spin–orbit torque in a magnetically doped topological insulator heterostructure. *Nature Materials* **13**, 699–704 (2014).

2. Mellnik, A. R. *et al.* Spin-transfer torque generated by a topological insulator. *Nature* **511**, 449–451 (2014).

3. Jamali, M. *et al.* Giant Spin Pumping and Inverse Spin Hall Effect in the Presence of Surface and Bulk Spin−Orbit Coupling of Topological Insulator Bi2Se3. *Nano Lett.* **15**, 7126–7132 (2015).

4. Katmis, F. *et al.* A high-temperature ferromagnetic topological insulating phase by proximity coupling. *Nature* **533**, 513–516 (2016).

5. Kane, C. L. & Mele, E. J. Quantum Spin Hall Effect in Graphene. *Phys. Rev. Lett.* **95**, 226801 (2005).

6. Bernevig, B. A., Hughes, T. L. & Zhang, S.-C. Quantum Spin Hall Effect and Topological Phase Transition in HgTe Quantum Wells. *Science* **314**, 1757–1761 (2006).

7. Wan, X., Turner, A. M., Vishwanath, A. & Savrasov, S. Y. Topological semimetal and Fermi-arc surface states in the electronic structure of pyrochlore iridates. *Phys. Rev. B* **83**, 205101 (2011).

8. Liu, D. F. *et al.* Magnetic Weyl semimetal phase in a Kagomé crystal. *Science* **365**, 1282–1285 (2019).

9. Belopolski, I. *et al.* Discovery of topological Weyl fermion lines and drumhead surface states in a room temperature magnet. *Science* **365**, 1278–1281 (2019).

10. Morali, N. *et al.* Fermi-arc diversity on surface terminations of the magnetic Weyl semimetal Co3Sn2S2. *Science* **365**, 1286–1291 (2019).

11. Zhang, S. S.-L., Burkov, A. A., Martin, I. & Heinonen, O. G. Spin-to-Charge Conversion in Magnetic Weyl Semimetals. *Phys. Rev. Lett.* **123**, 187201 (2019).

12. Burkov, A. A. Anomalous Hall Effect in Weyl Metals. *Phys. Rev. Lett.* **113**, 187202 (2014).

13. Wang, C., Gioia, L. & Burkov, A. A. Fractional Quantum Hall Effect in Weyl Semimetals. *Phys. Rev. Lett.* **124**, 096603 (2020).

14. Mazin, I. I. Robust half metallicity in $Fe_xCo_{1-x}S_2$. 4 (2000).

15. Shishidou, T., Freeman, A. J. & Asahi, R. Effect of GGA on the half-metallicity of the itinerant ferromagnet $CoS_2$. *Phys. Rev. B* **64**, 180401 (2001).

16. Leighton, C. *et al.* Composition controlled spin polarization in $Co_xFe_{1-x}S_2$ alloys. *J. Phys.: Condens. Matter* **19**, 315219 (2007).

17. Ramesha, K., Seshadri, R., Ederer, C., He, T. & Subramanian, M. A. Experimental and computational investigation of structure and magnetism in pyrite $Fe_xCo_{1-x}S_2$: Chemical bonding and half-metallicity. *Phys. Rev. B* **70**, 214409 (2004).

18. Wang, L. *et al.* $Co_{1-x}Fe_xS_2$: A Tunable Source of Highly Spin-Polarized Electrons. *Phys. Rev. Lett.* **94**, 056602 (2005).

19. Wang, L. *et al.* Composition controlled spin polarization in $Fe_xCo_{1-x}S_2$: Electronic, magnetic, and thermodynamic properties. *Phys. Rev. B* **73**, 144402 (2006).

20. Utfeld, C. *et al.* Bulk Spin Polarization of $Co_{(1-x)}Fe_xS_2$. *Phys. Rev. Lett.* **103**, 226403 (2009).

21. Teruya, A. *et al.* Large Cyclotron Mass and Large Ordered Moment in Ferromagnet $CoS_2$ Compared with Paramagnet $CoSe_2$. *J. Phys. Soc. Jpn.* **85**, 064716 (2016).

22. Wang, L., Chen, T. Y., Chien, C. L. & Leighton, C. Sulfur stoichiometry effects in highly spin polarized CoS2 single crystals. *Appl. Phys. Lett.* **88**, 232509 (2006).

23. Fujimori, A. *et al.* Resonant photoemission study of pyrite-type $NiS_2$, $CoS_2$ and $FeS_2$. *Phys. Rev. B* **54**, 16329–16332 (1996).

24. Wu, N. *et al.* The electronic band structure of CoS$_2$. *J. Phys.: Condens. Matter* **19**, 156224 (2007).

25. Wu, N. *et al.* The minority spin surface bands of CoS$_2$ (001). *J. Phys.: Condens. Matter* **21**, 295501 (2009).

26. Strocov, V. N. Intrinsic accuracy in 3-dimensional photoemission band mapping. *Journal of Electron Spectroscopy and Related Phenomena* **130**, 65–78 (2003).

27. Morris, B., Johnson, V. & Wold, A. Preparation and magnetic properties of cobalt disulfide. *Journal of Physics and Chemistry of Solids* **28**, 1565–1567 (1967).

28. Faber, M. S. *et al.* High-Performance Electrocatalysis Using Metallic Cobalt Pyrite (CoS2) Micro- and Nanostructures. *J. Am. Chem. Soc.* **136**, 10053–10061 (2014).

29. Schröter, N. B. M. *et al.* Chiral topological semimetal with multifold band crossings and long Fermi arcs. *Nat. Phys.* **15**, 759–765 (2019).

30. Schröter, N. B. M. *et al.* Observation and manipulation of maximal Chern numbers in the chiral topological semimetal PdGa. *arXiv:1907.08723 [cond-mat]* (2019).

31. Yang, Q. *et al.* Topological Engineering of Pt-Group-Metal-Based Chiral Crystals toward High-Efficiency Hydrogen Evolution Catalysts. *Advanced Materials* **32**, 1908518 (2020).

32. Li, G. & Felser, C. Heterogeneous catalysis at the surface of topological materials. *Appl. Phys. Lett.* **116**, 070501 (2020).

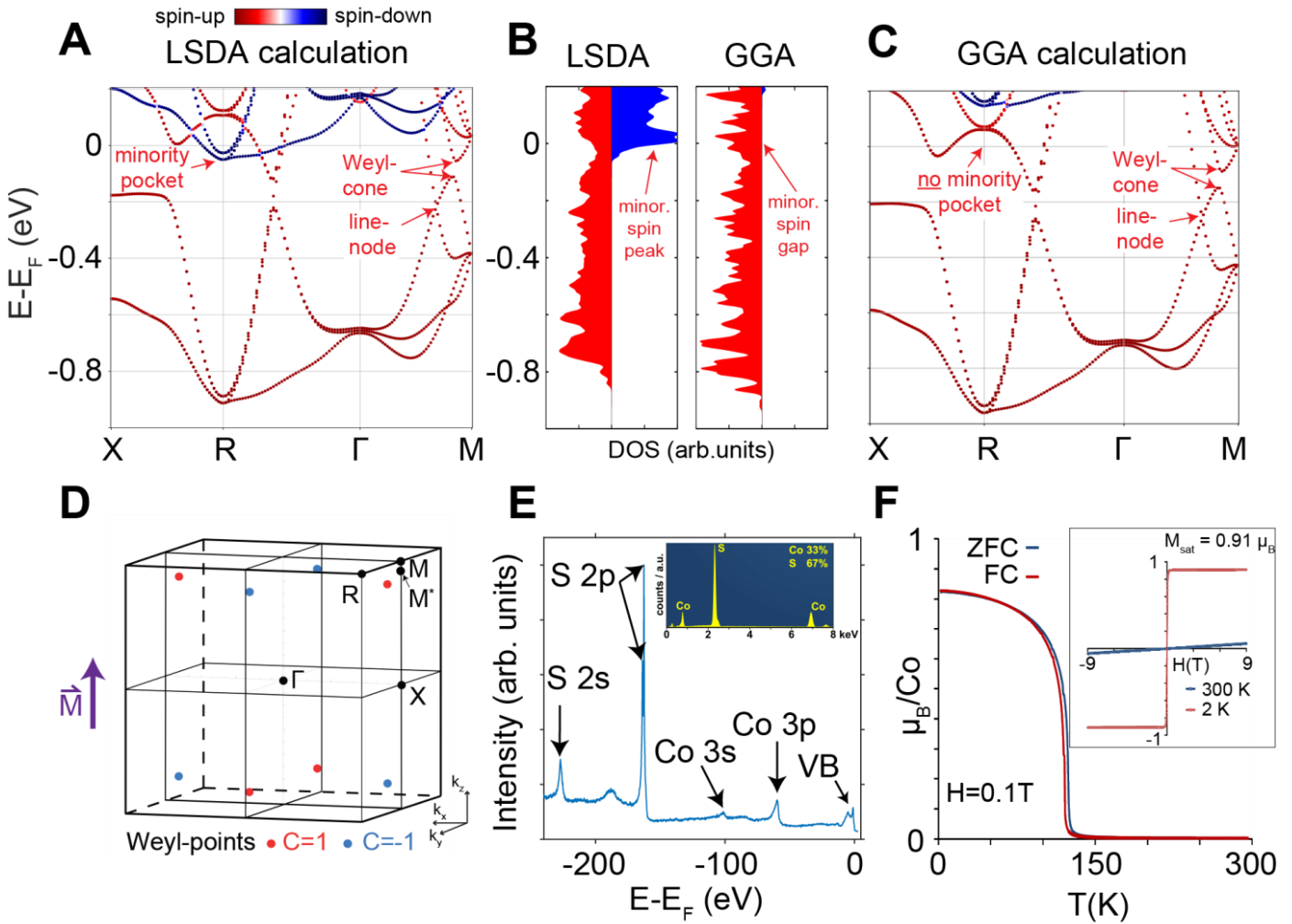

**Figure 1: Electronic structure and characterization of CoS$_2$ samples**
(A) Band structure obtained from local spin-density approximation (LSDA).
(B) Comparison of density of states (DOS) from both spin-channels for LSDA and generalized gradient approximation (GGA). The red solid area shows the majority spin, and the blue solid area the minority spins.
(C) Band structure obtained from GGA.
(D) Illustration of the Weyl-points in the bulk Brillouin zone, which are located on the high symmetry planes $k_x=0$ and $k_y=0$ parallel to the magnetisation direction $\vec{M}$. Red dots indicate Weyl points with positive Chern number C, blue dots indicate Wey points with negative Chern number.
(E) Core-level spectroscopy measured with photon energy hv=602 eV, arrows indicate elemental core levels and valence band (VB). Inset shows energy dispersive X-ray spectroscopy (EDX) curves, showing an ideal stoichiometry.
(F) Temperature dependence of magnetisation curve under zero field cooling (ZFC) as well as field cooled (FC) conditions, the applied field is 0.1T. The Curie point at 130 K is indicated. The inset shows the magnetic field dependence below and above the Curie temperature.

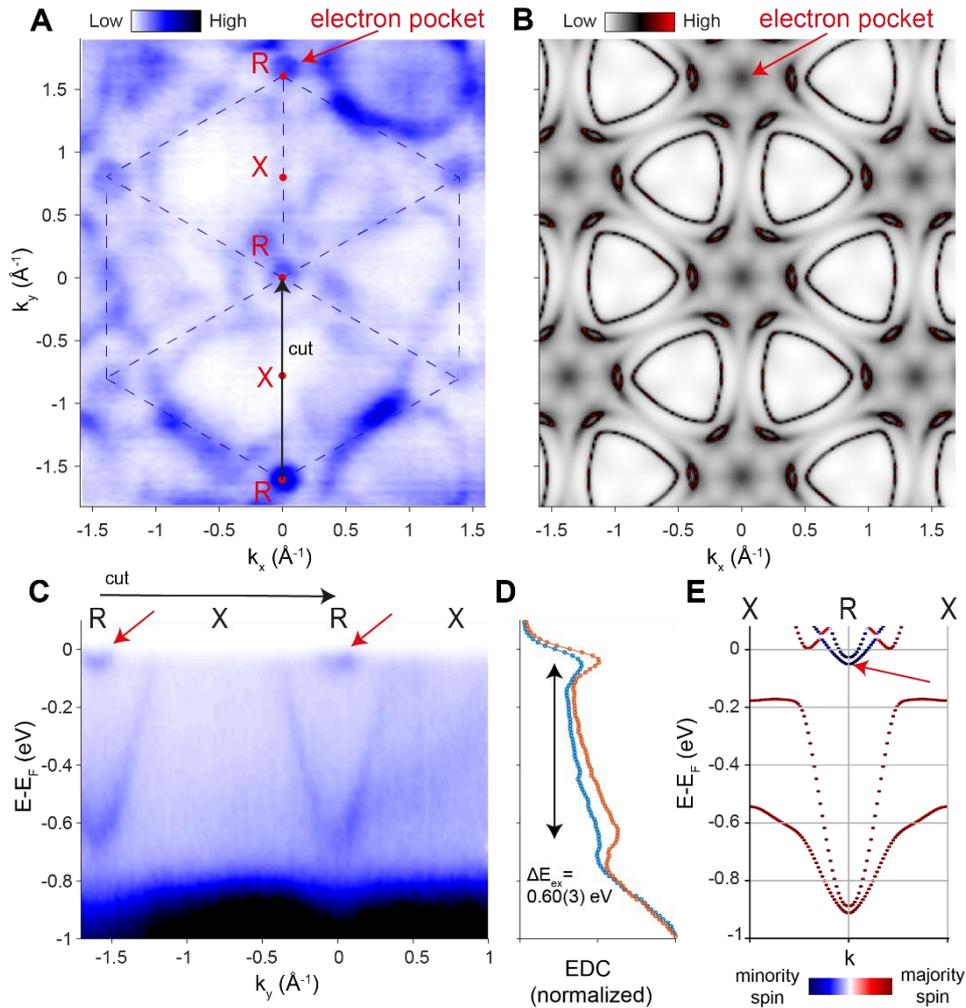

**Figure 2: Bulk band structure of the $k_z=\pi$ plane measured on the (111) surface.**

(A) Experimental Fermi-surface measured on the (111) cleavage plane with photon energy hv = 602 eV and linear-vertical polarization, integrated over 50 meV below the Fermi-energy. The red arrow indicates an electron pocket located at the R-point. The black arrow indicates the position of the line-cut shown in (C)

(B) Calculated Fermi-surface spectral function $A(k,E_F)$ for the same plane as shown in (A) obtained with the LSDA.

(C) Line-cut along the R-X-R direction as shown in (A), red arrows indicate the electron pockets at the R-point.

(D) Energy-distribution curves for the two R-points shown in (E). The black arrow indicates the magnitude of the exchange splitting of ΔE= 0.60(3) eV.

(E) Calculated band structure obtained with LSDA, red arrows indicate minority-spin electron pocket

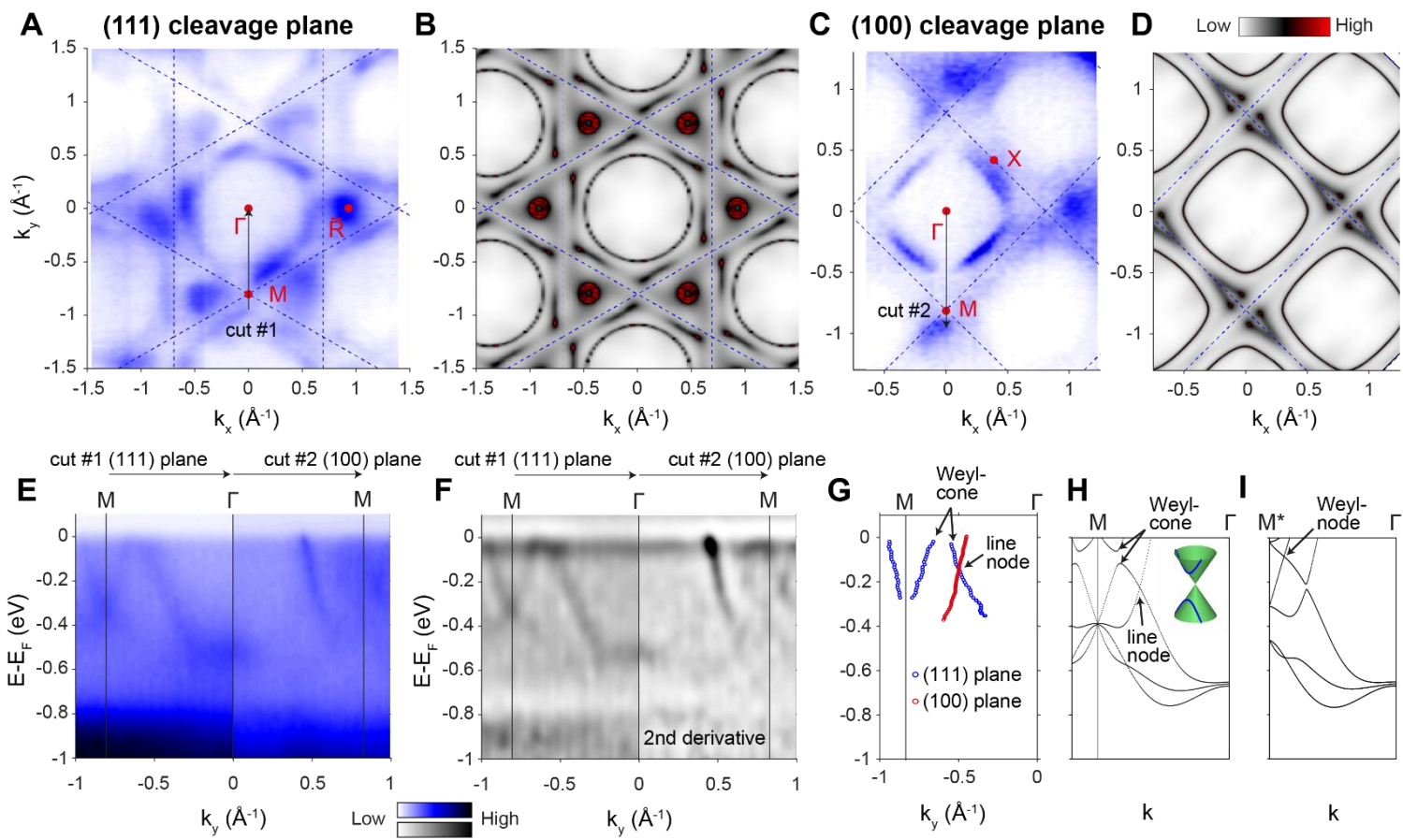

**Figure 3: Bulk band structure of the $k_z$=0 plane.**

(A) Experimental Fermi-surface measured on the (111) cleavage plane with photon energy hν = 512 eV and linear-vertical polarization, integrated over 50 meV below the Fermi-energy. The black arrow indicates the position of the line-cut shown in (E)

(B) Calculated Fermi-surface spectral function for the same plane as shown in (A) obtained with LSDA.

(C) Experimental Fermi-surface spectral function $A(k,E_F)$ measured on the (100) cleavage plane with photon energy hν = 475 eV, and linear-vertical polarization, integrated over 50 meV below the Fermi-energy. The black arrow indicates the position of the line cut shown in E.

(D) Calculated Fermi-surface for the same plane as shown in (C) obtained with LSDA.

(E) Line-cuts along the M-Γ direction from the (111) surface, as shown in (A), and the Γ-M direction from the (100) surface, as shown in (C).

(F) Second derivative spectrum of (E).

(G) Result of the momentum distribution curve (MDC) fitting of the bands along the M-Γ-M direction, where blue circles originate from data of the (111) plane, and red circles originate from data of the (100) plane.

(H) Calculated band dispersion along the M-Γ direction.

(I) Band dispersion along the M*-Γ direction where M* $(0,0.5,0.4581)\frac{2\pi}{a}$ is a point slightly displaced from the M point (see Fig. 1D), such that the k-path is passing through the Weyl point in the vicinity of M.

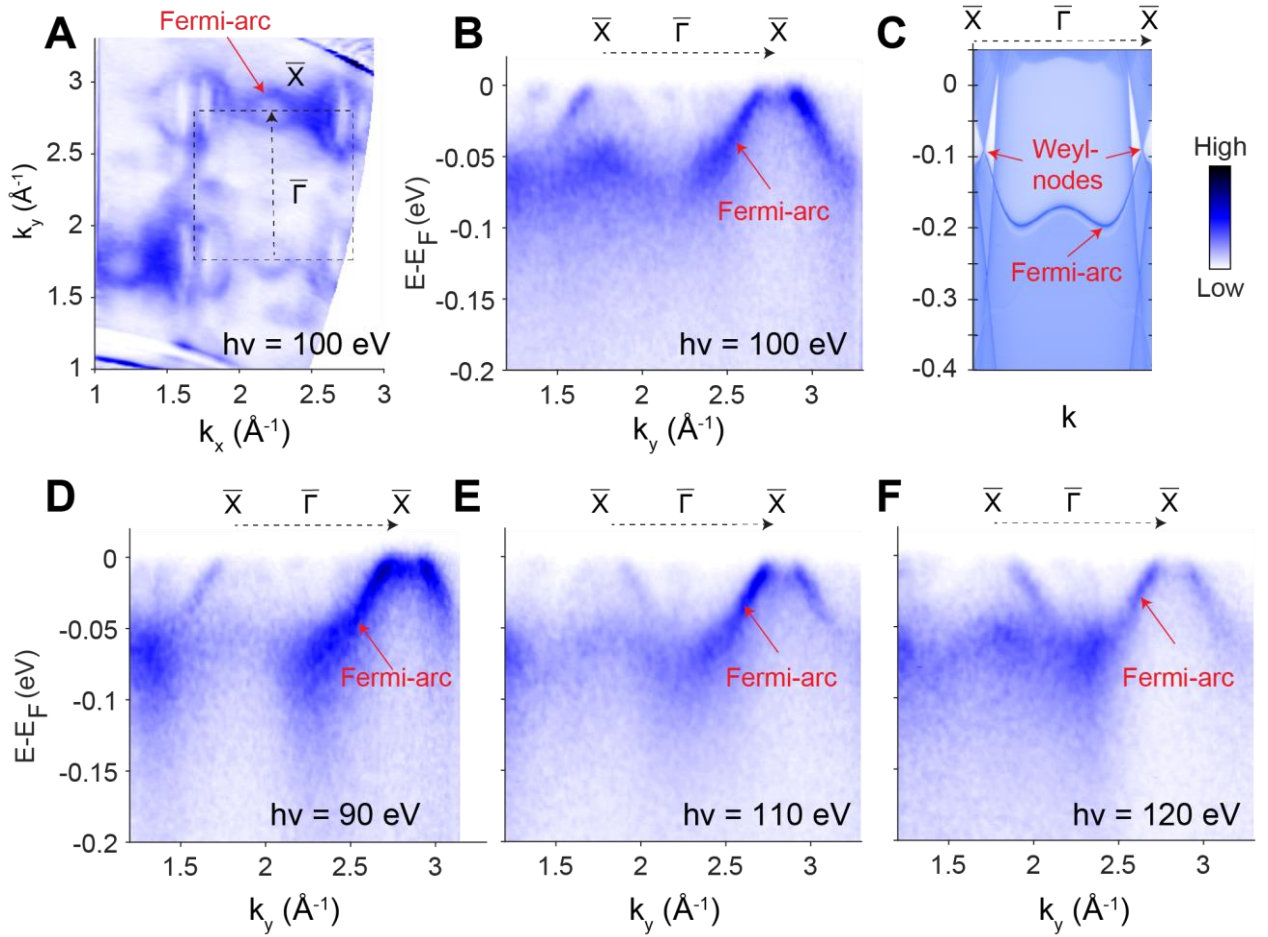

**Figure 4: Surface state structure on the (100) surface.**
(A) Experimental Fermi-surface measured on the (100) cleavage plane with photon energy hv = 100 eV and linear-horizontal polarization, integrated over 5 meV below the Fermi-energy. The dashed black line indicates the boundary of the surface Brillouin zone, and the black dashed arrow indicates the momentum direction of the line cuts shown in (B-F)
(B) Band dispersion of the Fermi-arc surface states, which are indicated by an red arrow. The photon energy used here was 100 eV and the polarization linear-horizontal.
(C) Calculated surface state dispersion along the same momentum direction as the experimental line-cuts.
(E-F) Same as in (B), but for photon energies of 90, 110, and 120 eV.

**Methods and supplementary materials for the manuscript "Observation of minority-spin carriers, topological surface states, and band folding in $CoS_2$"**

# Methods

## Sample growth

Single crystals of $CoS_2$ were synthesized using chemical vapor transport. The elements cobalt (200 mg, 1 eq., 99.5% Sigma Aldrich) and sulfur (217 mg, 2 eq., 99.5% Alfa Aesar) were mixed with 50 mg iodine (99.999% Sigma Aldrich) and sealed in an evacuated quartz glass ampoule. The ampoule was slowly (within 12 h) heated to 1000 °C kept there for 140h. The growth was along a natural temperature gradient of a tube furnace. An increase in crystal size can be accomplished using an additional slow cooling step from 1000 °C to 800 °C in 100h.

## ARPES measurements

Soft X-ray ARPES (SX-ARPES) measurements were performed at the SX-ARPES endstation [1] of the ADRESS beamline [2] at the Swiss Light Source, Switzerland, with a SPECS analyzer with an angular resolution of 0.1°. The photon energy varied from 350-1000 eV and the combined energy resolution was ranging between 50 meV to 150 meV. The temperature during sample cleaving and measurements was below 20 K and the pressure better than $1\times10^{-10}$ mbars. The increase of the photoelectron mean free path in the soft-X-ray energy range results, by the Heisenberg uncertainty principle, in a higher $k_z$ resolution of the ARPES experiment compared to measurements at lower photon energies [3], which was critical to measure the bulk band structure of $CoS_2$.

VUV-ARPES measurements were performed at the high-resolution ARPES branch line of the beamline I05 at the Diamond Light Source, UK [4]. Measurements at the high-resolution branch were performed with a Scienta R4000 analyzer, and a photon energy range between 90 eV and 130 eV, at a temperature below 20 K. Measurements in the VUV-ARPES regime are more surface sensitive than SX-ARPES and therefore most suitable to image the Fermi-arcs in $CoS_2$.

## Ab-initio calculations

We employed density functional theory (DFT) as implemented in the Vienna Ab Initio Simulation Package (VASP). For the Generalized Gradient Approximation (GGA) calculations, the exchange correlation term is described according to the Perdew-Burke-Ernzerhof (PBE) prescription together with projected augmented-wave (PAW) pseudopotentials, while of the Local Spin Density Approximation (LDA), the exchange correlation term is described according to the Dudarev simplified rotationally invariant approach, together with PAW pseudopotentials. The kinetic energy cut off was set to 400 eV. For the self-consistent calculation, a grid of 7x7x7 k-points was used. For DOS calculation, a grid of 11x11x11 k-points was used, with 1000 energy points. The spectral function of the Fermi-surfaces shown in Figs. 2-3 were computed with the programs Wannier90 [5] and WannierTools [6].

## Data availability

All data that supports the figures in this manuscript are available from the corresponding authors upon reasonable request.

[1] Strocov, V. N. et al. J. Synchrotron Radiat. 21, 32–44 (2014).
[2] Strocov, V. N. et al. J. Synchrotron Radiat. 17, 631–643 (2010).
[3] Strocov, V. N., Phys. Rev. Lett. 109 (2012) 086401.
[4] Hoesch, M. et al., Rev. Sci. Instrum. 88, 13106 (2017).
[5] G. Pizzi et al., J. Phys. Cond. Matt. 32, 165902 (2020).
[6] Wu, Q. et al., Computer Physics Communications. 224, 405–416 (2018).

# Supplementary materials
1. Sample characterization
2. Photon energy dependent ARPES measurements
3. Complementary experimental data of the bulk dispersion around the R-point
4. Effect of the magnetization direction on the Fermi-arc surface states
5. Further theoretical investigation of the nodal lines in $CoS_2$
6. Orbital character of the bulk band structure

## 1. Sample characterization

The samples were characterized using X-ray diffraction (XRD) and energy-dispersive X-ray spectroscopy (EDX). For structural analysis representative, smaller, crystals were ground. The diffraction pattern (Fig S1) was taken using a Stoe STADI-P powder diffractometer (Mo-Kα1 radiation, GE-monochromator). The lattice parameter was determined to be 5.5287(5) Å using the program WinXPOW Index v.3.5.0.2 and agrees well with the value from literature (5.539 Å [1]). The chemical composition was confirmed to be 1:2 of Co and S using a XL30 FEG-SEM with an EVEX EDX detector (see Fig 1E). A side phase of $Co_3S_4$ can be found in the diffraction pattern, however, the EDX analysis on larger single crystals did not show the same impurity. We attribute the side phase to small crystals which can be present in the reaction mixture and appear in powder patterns if ground. The impurity should neither affect our ARPES analysis or magnetic measurements that were taken on large single crystals.

[1] Brown, P. J., et al. "Magnetization distribution in CoS2; is it a half metallic ferromagnet?." *Journal of Physics: Condensed Matter* 17.10 (2005): 1583.

## 2. Photon energy dependent ARPES measurements

To determine the photon energy that corresponds to the high symmetry planes of the bulk band structure along the kz direction (perpendicular to the sample surface), we performed photon energy dependent measurements, which are displayed in Fig. S2. These photon energies were then used to perform the Fermi-surface maps in Figs. 2-3. To generate Fig. S2B, we used the free electron final state approximation

$$k_y = \sqrt{2mE_{kin}/\hbar} \sin\theta, \quad (S1)$$

$$k_z = \sqrt{\frac{2m(E_{kin} + V)}{\hbar^2} - k_y^2}, \quad (S2)$$

where θ is the emission angle, $E_{kin}$ the kinetic energy of the electrons (which is proportional to the photon energy), m the electron mass, and ℏ the reduced Planck constant.

To investigate the surface state character of the Fermi-arcs, we compared multiple Fermi surfaces of the 001 surfaces measured at low photon energies (hv=90-120 eV), which are displayed in Fig. S3. Since there is no noticeable dispersion with photon energy (i.e. when probing different $k_z$ momenta), we conclude that most bands present at the Fermi-surface are surface states or surface-resonance states, especially the Fermi-arcs that form circular hole pockets.

## 3. Complementary experimental data of the bulk dispersion around the R-point

Beyond the data measured on the (111) surface displayed in Fig. 2, we also obtained complementary information about the electron pockets at the R point from the (001) surface, which we show in Fig. S4. Fig. S4A displays the experimental Fermi-surface, and Fig. S4B the band dispersion along the R-X-R direction. We can clearly see the electron-like pocket centred at the R-point, which is also shown in Fig. 2 for the sample cleaved on the (111) direction. Note that we can also observe some weak signal at the Fermi-level in-between the R and X point, which may be due to another electron pocket. By comparison with the LSDA calculations shown in Fig. S4C, we observe that this additional electron pocket may be of majority-spin character.

## 4. Effect of the magnetization direction on the Fermi-arc surface states.

Our ab-initio calculations show no significant difference between the surface state structure for different surface planes of interest (e.g. (100), (010), and (001) surfaces for the same magnetization direction (001)), so we do not expect that the magnetisation direction plays a role in our experiment. Such an insensitivity of the surface state structure to the magnetisation direction can be expected, because the relevant energy scale that breaks the symmetries of the cubic crystal lattice is spin-orbit coupling, which is small in $CoS_2$ due to its relatively light elements. However, the direction of the magnetisation can affect the topological protection of the Fermi-arcs. For those domains for which the magnetisation is in the surface plane, the projection of the bulk pockets of the Weyl-cones to the surface Brillouin zone will carry either a Chern number of magnitude $|C|=1$ or $|C|=2$, so the Fermi-arcs must connect those projected bulk pockets due to the bulk-boundary correspondence. For any domains with out-of-plane magnetisation, projections of Weyl-points with opposite Chern numbers cancel (see Fig. 1D), so sufficiently strong perturbations may detach the Fermi-arcs from the projected bulk pockets, similar to Fermi-arcs in Dirac-semimetals [2]. Hence, polarising the magnetic domain structure with an external field could control the Fermi-arc structure at surfaces and interfaces.

[2] M. Kargarian, M. Randeria, Y.-M. Lu, Are the surface Fermi arcs in Dirac semimetals topologically protected? PNAS. **113**, 8648–8652 (2016).

## 5. Further theoretical investigation of the topological nodal-lines in $CoS_2$

We computed the gap between the two last valence bands in the $k_z=0$ plane, where there is a mirror symmetry that can protect the crossing of bands. We found that some bands cross in that plane, with the crossing points being connected in a nodal line, as depicted in Figure S5.

Apart from the non-symmorphic symmetry enforced degeneracy at the boundaries of the BZ, one can observe two branches of nodal line coming from $k_x=\pi$ plane to $k_x=-\pi$. Following the same argument, the mirror can also protect crossings in the $k_z=\pi$ plane. When taking into account magnetism only, the subspaces of spin up and spin down still preserve the full symmetry of the space group, such that bands are degenerate in the $k_z=\pi$ plane. However, when SOC is introduced, the symmetry breaking takes effect, where the energy scale of the symmetry breaking is of the order of the SOC. The degeneracy is lifted on the $k_z=\pi$ plane everywhere except for some nodal lines. This degeneracy lifting, however, remains of the order of the SOC, which is small in $CoS_2$, so that the numerical determination of the nodal line in the $k_z=\pi$ plane is not as sharp as in the $k_z=0$ plane due to the finite energy resolution of our ab-initio calculations (see Fig. S5).

To check the topological nature of the nodal lines, one can compute the Berry phase in the $k_z$ direction for several values of $k_x$ and $k_y$. One particular useful way to do this is to compute Wilson Loops (WL), integrating along $k_z$ direction for lines in the ($k_x$, $k_y$) plane. We performed the computation in two particular directions as showed in Figure S6, where it can be seen the path of the WL. Whenever the Berry phase is computed in a ($k_x$,$k_y$) point that lays inside one of the nodal lines, the result is π, signalling that along that direction there should be drumhead states in the (001) surface if one preserves the unit cell. When the ($k_x$,$k_y$) points lays outside of the nodal lines or inside the projection of two, then the Berry phase is 0, signalling that the possible surface states that are in that direction are not topologically protected.

## 6. Orbital character of the bulk band structure

The orbital weights of the band structure of $CoS_2$ close to the Fermi-level is shown in Fig. S7. We can see that all bulk bands are of Cobalt d-electron character, which means that the Fermi-arcs that are derived from these bands will also be of d-electron character.

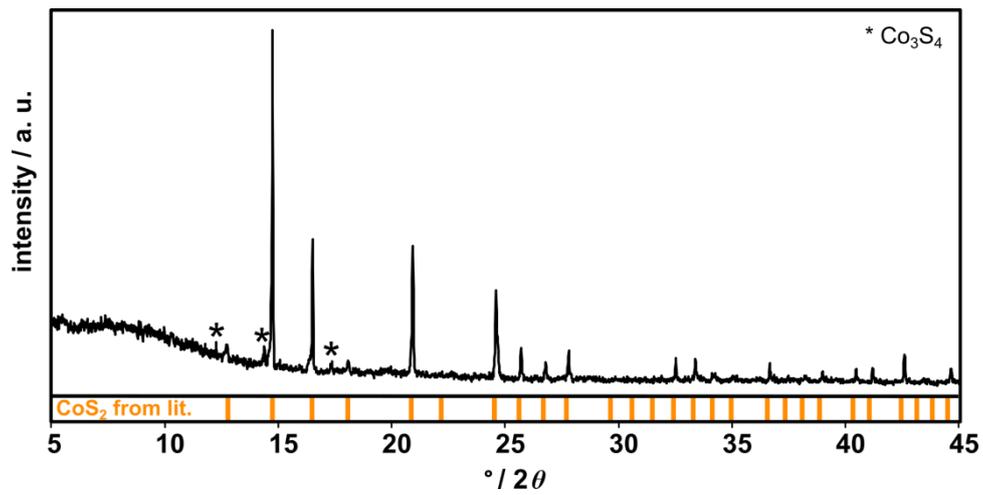

**Fig S1: X-ray diffraction pattern of ground CoS$_2$ crystals compared to the reflex positions based on Brown et al. (orange).[1]** Main peaks of a side phase (Co$_3$S$_4$) are marked with an asterisk.

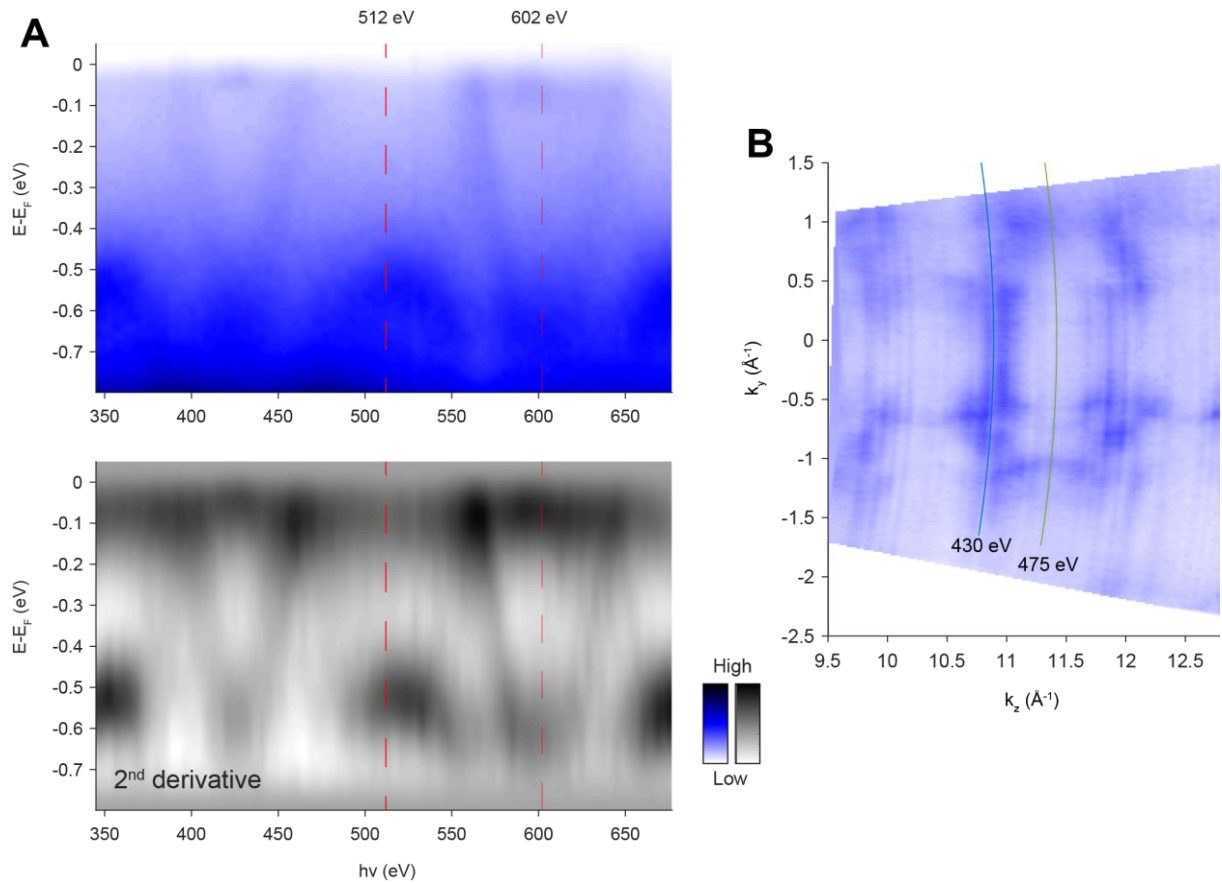

**Fig S2: Photon energy dependent SX-ARPES measurements of bulk band structure**
**(A) Top:** Raw ARPES data showing the dispersion along the $k_z$ direction for the sample cleaved on the (111) plane. **Bottom:** 2$^{nd}$ derivative spectrum of the data shown on the top. Red dashed lines indicate high symmetry points at Γ ($k_z=0$, 512 eV) and R ($k_z=\pi$, 602 eV).
**(B)** Raw ARPES data showing $k_z$ vs $k_y$ Fermi-surface. For the conversion to $k_z$ coordinates, the free electron final state approximation with inner potential of $V_0=27$ eV was used. The green and blue lines indicate the high symmetry planes at $k_z=0$ (475 eV) and $k_z=\pi$ (430 eV), respectively.

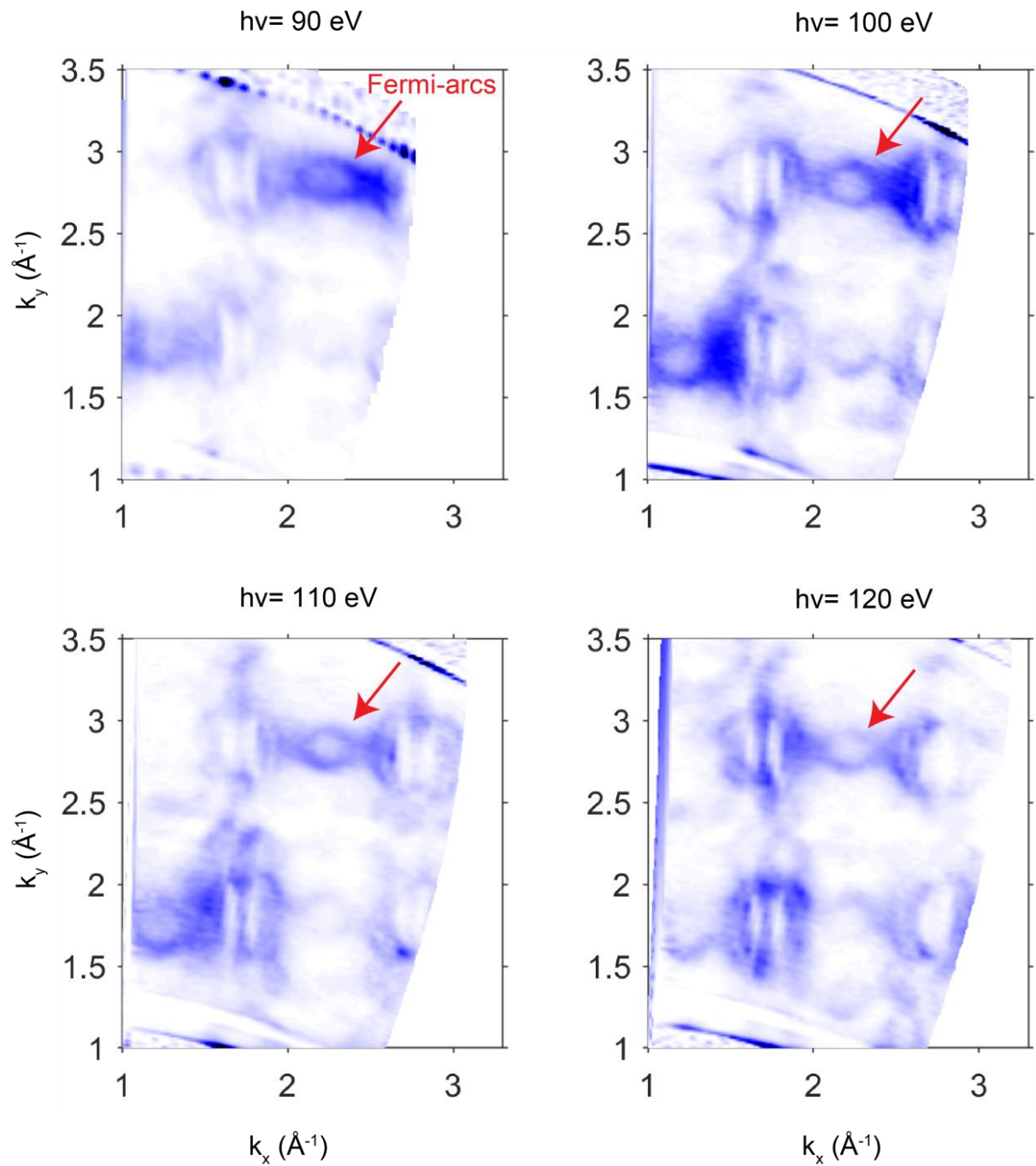

**Fig S3: Photon energy dependent VUV-ARPES measurements of surface band structure**
Fermi-surfaces measured between 90-120 eV photon energy. Notice that there are some minor distortions because the measurements were performed on a strongly tilted surfaces, which may have caused a loss of ideal focusing conditions when the sample was rotated during the Fermi-surface mapping.

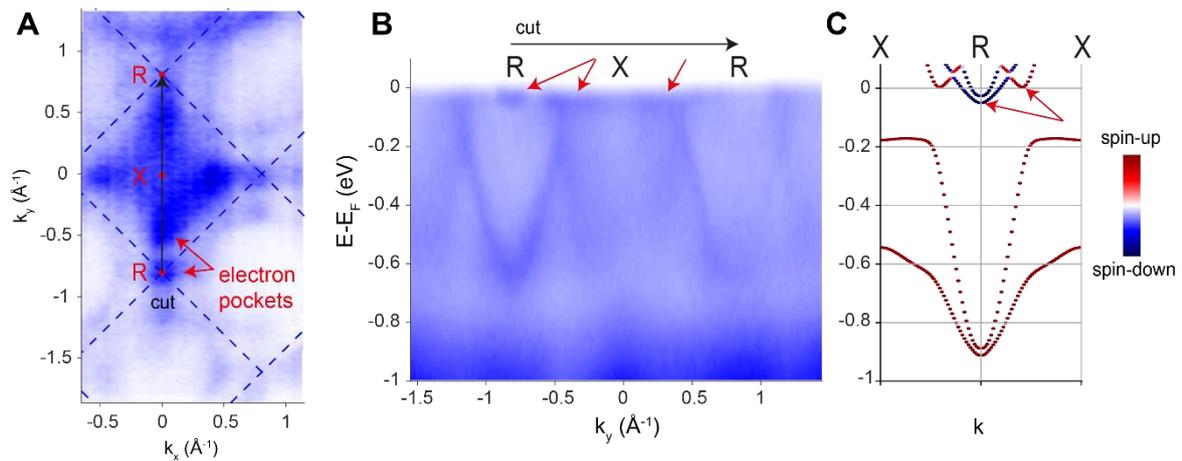

**Fig S4: Complementary measurements of the $k_z=\pi$ plane from the (001) surface**

(A) Experimental Fermi-surface measured with hv=430 eV photon energy. Red arrows indicate electron pockets and the black arrow indicates the line-cut shown in (B).

(B) Band-dispersion along the R-X-R direction. Red arrows indicate electron pockets

(C) Calculated band dispersion obtained with the LSDA calculation, red arrows indicate electron pockets observed in the experimental data.

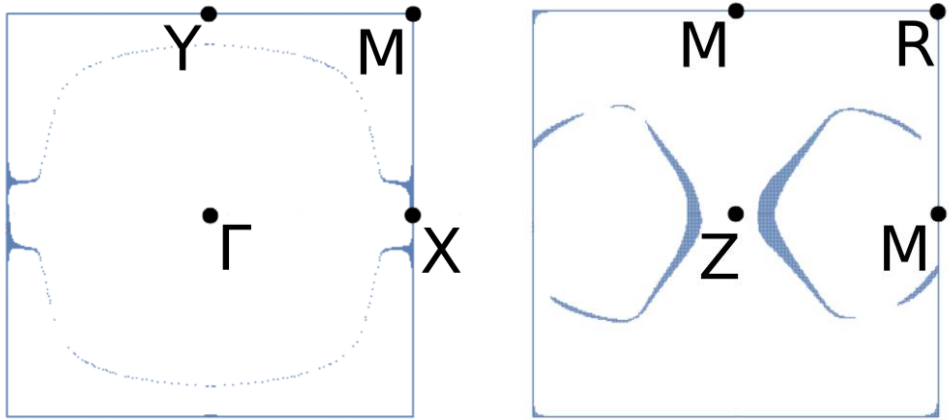

**Fig S5: Ab-initio calculation of the nodal-line dispersions in CoS$_2$**
(left) line-nodes in the kz=0 plane
(right) line-nodes in the kz=π plane

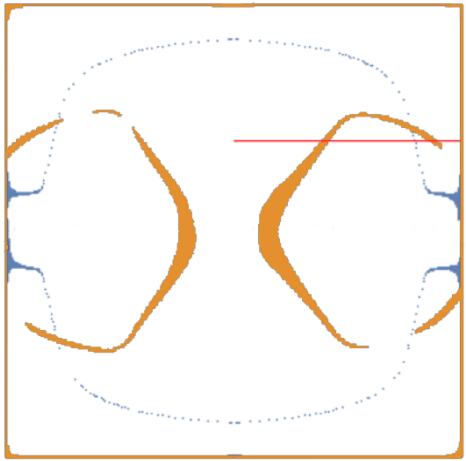 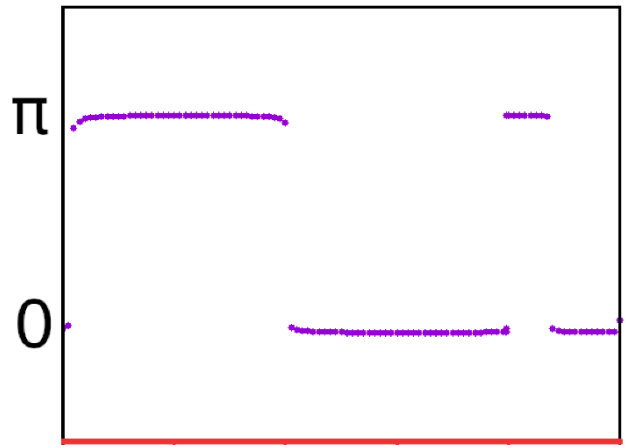

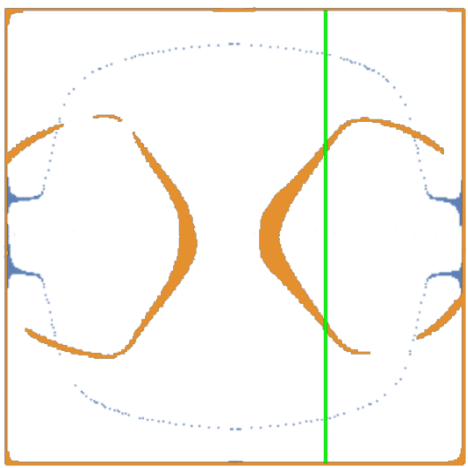 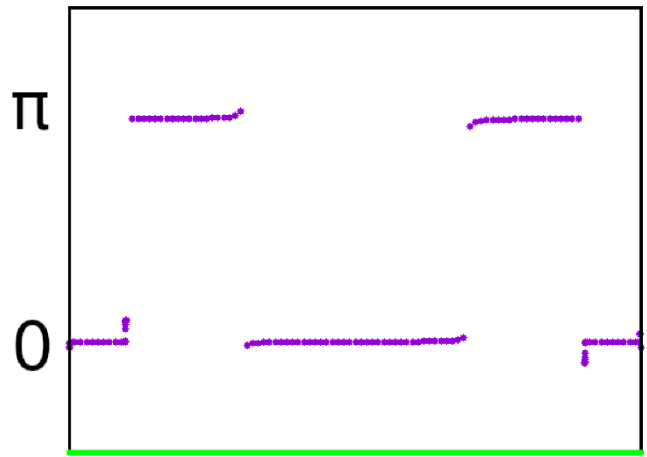

**Fig S6: Wilson loop calculations along different directions in the Brillouin zone**

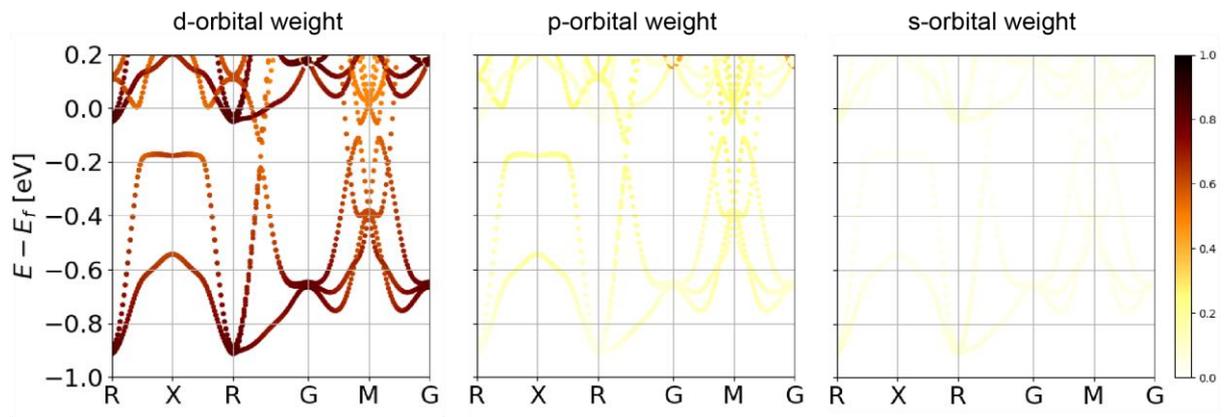

**Fig S7: Orbital weights of the bands close to the Fermi-level**